\documentclass[aps,prc,groupedaddress,showpacs,showkeys]{revtex4}
\usepackage{latexsym,epsfig,amsmath,amssymb}
\usepackage{graphicx}
\usepackage{graphics}

\newcommand{\seq}{\begin{subequations}}
\newcommand{\sen}{\end{subequations}}
\newcommand{\eq}{\begin{eqnarray}}
\newcommand{\en}{\end{eqnarray}}
\newcommand{\ra}{\rangle}
\newcommand{\la}{\langle}

\def\L2{\Lambda^2}

\begin{document}

\title{Phenomenological Lagrangian approach \\ 
       to the electromagnetic deuteron form factors}
\author{Yubing Dong$^{1,2}$, 
Amand Faessler$^{3}$, 
Thomas Gutsche$^{3}$, 
Valery E. Lyubovitskij$^{3}$\footnote{On leave of absence
        from Department of Physics, Tomsk State University,
        634050 Tomsk, Russia}
\vspace*{1.2\baselineskip}}
\affiliation{
$^{1}$Institute of High Energy Physics, Chinese Academy of Sciences, 
Beijing 100049, P. R. China
\vspace*{1.2\baselineskip} \\
$^{2}$Theoretical Physics Center for Science Facilities (TPCSF), 
CAS, P. R. China 
\vspace*{1.2\baselineskip} \\ 
$^{3}$Institut f\"ur Theoretische Physik,
Universit\"at T\"ubingen,\\
Auf der Morgenstelle 14, D--72076 T\"ubingen, Germany\\} 

\date{\today}

\begin{abstract}

A phenomenological Lagrangian approach is employed to study 
the electromagnetic properties of the deuteron. The deuteron is 
regarded as a weakly bound state of the proton and neutron. 
We construct a general form for the electromagnetic one-- and two--body
transition operators formulated in terms of the nucleon fields, which  
are then used in the calculation of the electromagnetic deuteron 
form factors. One of the two--body operators is 
responsible for explaining the quadrupole 
moment form factor. We show that in our approach the data 
on the deuteron form factors as well as on the differential cross 
section of elastic electron--deuteron scattering are well explained.     

\end{abstract}



\maketitle

\newpage

\section{Introduction}

The study of the electromagnetic properties of the deuteron has a long 
and rich history (for some recent reviews, see,  
e.g.~\cite{Garcon:2001sz}-\cite{Gross:2002ge}). 
The deuteron, as a spin--1 particle, is usually believed 
to be a weakly bound system of a proton and a neutron 
with a binding energy $\epsilon_D \sim 2.22$~MeV. 
Since the electromagnetic (EM) 
properties of the deuteron can also shed light on the EM form factors of the 
neutron as well as on nuclear effects on the form factors, the study of the 
deuteron form factors with a EM probe is of great interest. 

The matrix element for elastic electron--deuteron ($eD$) 
scattering in the one--photon approximation is  
\eq 
{\cal M}=\frac{e^2}{Q^2} \bar u_e(k^\prime) \gamma_{\mu} 
u_e(k) {\cal J}_{\mu}^D(p,p^\prime)
\en 
where $k$ and $k^\prime$ are the four--momenta of initial and final 
electrons and ${\cal J}_{\mu}^D(p,p^\prime)$ is the deuteron EM current -- 
\eq\label{D_current} 
{\cal J}_{\mu}^D(p,p^\prime) = 
- \biggl( G_1(Q^2)\epsilon^{\prime *}\cdot\epsilon-\frac{G_3(Q^2)}{2m_D^2}
\epsilon\cdot q\epsilon^{\prime *}\cdot q \biggr) (p + p^\prime)_{\mu} 
- G_2(Q^2) \biggl( \epsilon_{\mu}\epsilon^{\prime *}\cdot q 
- \epsilon^{\prime *}_{\mu} \epsilon\cdot q \biggr)~, 
\en 
where $m_D$ is the deuteron mass, $\epsilon$($\epsilon^\prime$) and 
$p(p^\prime)$ are polarization and four--momentum of the initial (final) 
deuteron with $q=p^\prime - p$ being the momentum transfer.
The three EM form factors $G_{1,2,3}$ of the deuteron 
are related to the charge $G_C$, quadrupole $G_Q$ and magnetic $G_M$ 
form factors by
\eq 
G_C = G_1+\frac23\tau G_Q\,, \hspace*{.25cm} 
G_M \ = \ G_2 \,,            \hspace*{.25cm}
G_Q = G_1-G_2+(1+\tau)G_3,   \hspace*{.25cm} 
\tau=\frac{Q^2}{4m_D^2} \,.
\en 
These form factors are normalized at zero recoil as 
\eq 
G_C(0)=1\,, \ \ \ 
G_Q(0)=m_D^2{\cal Q}_D=25.83\,, \ \ \
G_M(0)=\frac{m_D}{m_N}\mu_D=1.714 \, ,
\en 
where $m_N$ is the nucleon mass, 
${\cal Q}_D$ and $\mu_D$ are the quadrupole and magnetic moments 
of the deuteron. 
Since the deuteron is a spin--1 particle it has three EM form factors 
in the one--photon--exchange (OPE) approximation, due to current conservation
and the $P$ and $C$ invariance of the EM interaction. The three  
form factors $G_{E,M,Q}$ can be determined by measuring the unpolarized,
elastic $eD$ differential cross sections and one of polarization observables, 
like the deuteron polarization tensor 
\eq 
T_{20} = - \frac{1}{{\cal S}\sqrt{2}} 
\biggl( \frac{8}{3} \tau G_C G_Q + \frac{8}{9} \tau^2 G_Q^2 
+ \frac{\tau}{3} \biggl( 1 + 2 (1 + \tau) {\rm tan}^2\frac{\theta_e}{2} \biggr)
G_M^2 \biggr)
\en 
with 
${\cal S} = A(Q^2)+B(Q^2)\tan^2(\theta_e/2)$. 
The two form factors $A(Q^2)$ and $B(Q^2)$ are related to the EM 
form factors of the deuteron as 
\seq 
\eq 
A = G_C^2+\frac{2}{3}\tau G_M^2+\frac89\tau^2G_Q^2\,,\hspace*{.5cm} 
B = \frac{4}{3} \tau \, (1 + \tau) G_M^2 \,. 
\en 
\sen 
According to the Rosenbluth separation~\cite{Rosenbluth:1950yq}, the 
elastic scattering of an unpolarized electron from the deuteron results 
in an ${\cal O}(\alpha^2)$ differential cross section~\cite{Arnold:1979cg}  
$d\sigma/d\Omega = (d\sigma/d\Omega)_{\rm Mott} \ {\cal S}$, 
where $\theta_e$ is the electron scattering angle in the 
laboratory frame of the collision and $(d\sigma/d\Omega)_{\rm Mott}$ 
is the Mott cross section. 

The theoretical study of $eD$ elastic scattering and deuteron EM form factors 
has been performed in different 
approaches~\cite{theory1}-\cite{Ivanov:1995zp}: 
potential models, phenomenological models including quark, meson and 
nucleon degrees of freedom, effective field theories, etc. (for an overview 
see~\cite{Garcon:2001sz}-\cite{Gross:2002ge}).  
In the present work we apply a phenomenological Lagrangian approach to 
study the EM form factors of the deuteron. We consider the deuteron as  
a weakly bound system of proton and neutron. The coupling of the 
deuteron to its constituents is determined by the compositeness 
condition $Z=0$~\cite{Weinberg:1962hj,Efimov:1993ei},
which implies that the renormalization constant of the hadron
wave function is set equal to zero.
Note, that this condition was originally also applied to the study of
the deuteron as a bound state of proton and neutron~\cite{Weinberg:1962hj}.
Then it was extensively used in low--energy hadron
phenomenology as the master equation for the treatment of
mesons and baryons as bound states of light and heavy
constituent quarks (see Refs.~\cite{Ivanov:1996pz}). 
In Refs.~\cite{Faessler:2007gv} this condition was used in the 
application to hadronic molecule configurations of light and heavy mesons. 
 
To study the EM form factors of the deuteron, in a first step we employ 
the empirical EM couplings of the photon to the proton (or neutron) in
the one--body operators to set up the well determined part
of the photon-deuteron coupling. As a consequence of the non--local
description of the deuteron bound state direct contact terms are then included
to guarantee local gauge invariance of the EM interaction.   
In the last step we introduce additional, phenomenological two-body operators,
which are assumed to represent e.g. meson--exchange currents and 
in turn imply a $D$--wave component in the deuteron wave function.
Parameters of two--body operators are deduced from a fit to the 
available data. 

The paper is organized as follows. In Sec. II, we discuss the basic 
notions of our approach: the coupling of the deuteron to its constituents
involving the compositeness condition and the derivation of the EM one-- and 
two--body operators contributing to the form factors of the deuteron. 
In Sec. III we discuss the numerical results for the deuteron 
EM form factors as well as the two form factors $A(Q^2)$ and $B(Q^2)$
entering in 
the differential cross section of the $eD$ elastic scattering. 
In Sec. IV we give our conclusions. 

\section{Approach} 

\subsection{Deuteron as a proton--neutron bound state} 

In this section we discuss the formalism for the study of the
deuteron interpreted as a hadronic molecule -- a weakly bound state 
of proton and neutron: $|D\ra = |pn\ra$. 
We write the deuteron mass $m_D$ in the form $m_D = 2 m_N - \epsilon_D$, 
where $m_N = m_p = 0.93827$ GeV is the nucleon mass and
$\epsilon_D \simeq 2.22$ MeV is the binding energy.
Based on our approach, the coupling
of the deuteron to its two constituents -- proton and neutron, is  
\eq\label{Lagr_D} 
{\cal L}_D(x) = g_{D} D^\dagger_\mu(x)\int dy 
\Phi_D(y^2) p(x + y/2) C \gamma^{\mu}
n(x - y/2)+ {\rm H.c.},
\en 
where $C = \gamma^0 \gamma^2$ is the charge conjugation matrix and $x$ 
is the center--of--mass (CM) coordinate.
The correlation function $\Phi_D$ characterizes the finite size
of the deuteron as a $pn$ bound state and depends on the relative 
Jacobi coordinate $y$.
A basic 
requirement for the choice of an explicit form of the correlation 
function is that its Fourier transform vanishes sufficiently fast in 
the ultraviolet region of Euclidean space to render the Feynman 
diagrams ultraviolet finite. We adopt the Gaussian form, 
$\tilde\Phi_D(p_E^2) \doteq \exp( - p_E^2/\Lambda_D^2)\,,$
for the Fourier transform of the vertex function, where $p_{E}$ is 
the Euclidean Jacobi momentum. Here, $\Lambda_D$ is a size parameter  
that characterizes the distribution of the constituents in the 
deuteron. 

The coupling constant $g_D$ in Eq.~(\ref{Lagr_D}) is determined by the 
compositeness condition~\cite{Weinberg:1962hj,Efimov:1993ei},
which implies that the renormalization constant of the hadron
wave function is set equal to zero: 
\eq 
\label{ZX}
Z_D &=& 1 - \Sigma^\prime_D(m_D^2) = 0 \,.
\en 
Here, $\Sigma^\prime_D(m_{D}^2) = g_{_{D}}^2 \Pi^\prime_D(m_D^2)$ is the
derivative of the transverse part of the mass operator
$\Sigma^{\alpha\beta}_D$, conventionally split into the transverse
$\Sigma_D$ and longitudinal $\Sigma^L_D$  parts as:
\eq 
\Sigma^{\alpha\beta}_{D}(p) = g^{\alpha\beta}_\perp \Sigma_D(p^2) 
+ \frac{p^\alpha p^\beta}{p^2} \Sigma^L_D(p^2) \,,
\en 
where 
$g^{\alpha\beta}_\perp = g^{\alpha\beta} - p^\alpha p^\beta/p^2\,,
\hspace*{.2cm} g^{\alpha\beta}_\perp p_\alpha = 0\,.$ 
The mass operator of the deuteron is described in Fig.1. 
For a fixed value of the size parameter $\Lambda_D$ the coupling 
$g_D$ is determined according to the compositeness condition. 

To clarify the physical meaning of the compositeness condition, 
we reiterate that the renormalization constant $Z_D^{1/2}$ can also be 
interpreted as the matrix element between the physical and the 
corresponding bare states. For $Z_D=0$ it then follows that the physical state
does not contain the bare one and hence the deuteron is described as a 
bound state of the proton and neutron. As a result of the interaction of the 
deuteron with its constituents, the deuteron is dressed, i.e. its mass and 
its wave function have to be renormalized. 

Following Eq.~(\ref{ZX}) the coupling constant $g_D$ 
can be expressed in the form:
\eq 
\frac{1}{g_D^2}&=&\frac{1}{8\pi^2} 
\int_0^{\infty} \int_0^{\infty} \frac{d\alpha d\beta}{(1+\alpha+\beta)^3}
\exp\biggl( - 2 (\alpha+\beta)\mu_N^2 
+ \frac{\alpha+\beta+4\alpha\beta}{2(1+\alpha+\beta)}\mu_D^2  
\biggr)\nonumber \\
&\times&\Bigg( (\alpha+\beta+4\alpha\beta)  
\biggl( \mu_N^2+\frac{1}{2 (1+\alpha+\beta)}
+\frac{(1+2\alpha)(1+2\beta)}{4(1+\alpha+\beta)^2}\mu_D^2 \biggr)  
+\frac{(1+2\alpha)(1+2\beta)}{2(1+\alpha+\beta)}\Bigg) \,, 
\en 
where $\mu_H = m_H/\Lambda_D$ with $H=N, D$. 
 
\subsection{Matrix element of the photon-deuteron interaction} 

To calculate the deuteron EM form factors, we construct 
the electromagnetic transition operator including one-- and two--body 
parts and formulated in terms of nucleon degrees of freedom -- the
constituents of the deuteron. 
Note, that the direct coupling of the deuteron with the photon field 
vanishes because the $Z_D$ factor equals zero. It guarantees that
double counting is avoided. 

We construct the one-- and two--body operators in a general phenomenological 
form, which principally includes all possible corrections (e.g. meson--cloud 
effects) and the dependence on the photon momentum (form factors). 
The one--body operator (or its 
Fourier transform) with $J_\mu^{(1)}(q) = J_\mu^{NN}(q) + J_\mu^{DNN}(q)$ 
contains two terms. The first term, $J_\mu^{NN}(q)$, is generated by 
the coupling of nucleons to the electromagnetic field: 
\eq 
J_\mu^{NN}(q) = \int d^4 x e^{-iqx}   
\bar N(x) \biggl( \gamma^{\mu} F_1^N(q^2) 
+ \frac{i\sigma^{\mu\nu}q_{\nu}}{2m_N}F^N_2(q^2) \biggr) N(x) \,, 
\en 
where $F_1^N(q^2)$ and $F_2^N(q^2)$ with $N=p,n$ are the conventional 
Dirac and Pauli form factors of the nucleon [see Fig.2(a)]. 
The second term, $J_\mu^{DNN}(q)$, is generated by gauging
the nonlocal strong Lagrangian ${\cal L}_D$ [see Fig.2(b)].  
To restore electromagnetic gauge invariance in 
${\cal L}_D$, the proton field should be multiplied 
by the gauge field exponential (see further details 
in~\cite{Ivanov:1996pz}): 
\eq\label{Subs_nonmin} 
p(y) \to e^{i e I(x,y,P)} p(y) \, , \hspace*{1cm} 
I(x,y,P) = \int\limits_y^x dz_\mu A^\mu(z).
\en 
For the derivative of $I(x,y,P)$ we use the
path--independent prescription suggested in~\cite{Mandelstam:1962mi}
which in turn states that the derivative of $I(x,y,P)$ does
not depend on the path $P$ originally used in the definition. 
The nonminimal substitution~(\ref{Subs_nonmin}) is therefore 
completely equivalent to the minimal prescription. Expanding the 
exponential $e^{i e I(x,y,P)}$ in powers of 
the electromagnetic field and keeping the linear term 
(corresponding to the vertex $D^\dagger p n \gamma 
+ {\rm H.c.}$) we generate an additional contribution [see Fig.2(b)]  
to the electromagnetic one--body operator: 
\eq 
J_\mu^{DNN}(q) = - ig_D \int d^4 x d^4 y  
D^\dagger_\nu(x) \Phi_D(y^2) p(x+y/2) C \gamma^\nu n(x-y/2) 
\int\limits_x^{x+y/2} dz_\mu e^{-iqz} + {\rm H.c.} 
\en 
There is a number of possible contributions to the two--body 
operator $J_\mu^{(2)}(q) = J_\mu^{4N}(q)$. We restrict to the three 
simplest terms with the smallest number of derivatives 
[see the general diagram of Fig.2(c)]: 
\seq\label{two_body} 
\eq 
J_\mu^{4N}(q) &=& J_\mu^{4N; 1}(q) + J_\mu^{4N; 2}(q) 
+ J_\mu^{4N; 3}(q) \,, \\
 J_\mu^{4N; 1}(q) &=& \int d^4 x e^{-iqx} g_1 F_1^{NN}(q^2) 
\bar n(x) \gamma^\alpha C \bar p(x) 
p(x) C \gamma_\alpha i \sigma_{\mu\nu} q^\nu n(x) + {\rm H.c.}\,, \\ 
 J_\mu^{4N; 2}(q) &=& \int d^4 x e^{-iqx} g_2 F_2^{NN}(q^2) 
\bar n(x) \!\!\not\! q \, C \bar p(x) 
p(x) C i \sigma_{\mu\nu} q^\nu n(x)  + {\rm H.c.}\,, \\ 
 J_\mu^{4N; 3}(q) &=& \int d^4 x e^{-iqx} g_3 F_3^{NN}(q^2) 
[\bar n(x) \gamma^\alpha  C \bar p(x)]  
i (\!\stackrel{\rightarrow}{\partial}_{\,\mu} 
-  \!\stackrel{\leftarrow}{\partial}_{\,\mu}) 
[p(x) C \gamma_\alpha  n(x)]\,, 
\en
\sen 
where 
$g_i$ and $F_i^{NN}(q^2)$ are the phenomenological electromagnetic 
two--body nucleon couplings and form factors, respectively. 

To calculate the EM form factors of the deuteron we project 
the dressed operator $J_\mu(q) = J_\mu^{(1)}(q) + J_\mu^{(2)}(q)$ 
between the deuteron states: 
\eq\label{D_matrix} 
\la D(p^\prime) | J_\mu(q) | D(p) \ra  = 
(2\pi)^4 \ \delta^4(p^\prime - p - q) \ 
{\cal J}_{\mu}^D(p,p^\prime) \; , 
\en 
where ${\cal J}_{\mu}^D(p,p^\prime)$ is the deuteron EM current given by 
the expression (\ref{D_current}). The diagrams contributing 
to the matrix element~(\ref{D_matrix}) are shown in Fig.3 -- 
the diagrams generated by the one--body currents [Figs.3(a)-3(c)]  
and by the two--body currents [Fig.3(d)]. To evaluate these  
diagrams we take the $T$--product of the EM current in the $S$--matrix 
defined in terms of the interaction Lagrangian ${\cal L}_D$ and 
use the standard free fermion propagators for the nucleons
in the loops. Let us stress that a similar approach to the one presented
here was previously developed 
in~\cite{Ivanov:1995zp}. However, in Ref.~\cite{Ivanov:1995zp} 
the authors did not consider two-body operators. Also note that the    
deuteron EM current generated by both the one--body and two--body 
nucleon operators is manifestly Lorentz covariant and gauge 
invariant. 

We want to point out again that our approach is purely phenomenological 
in the sense that we do not calculate the one- and two--body operators 
from microscopic models, but constrain their forms using a justified 
physical background: in particular, the two--body operators~(\ref{two_body}) 
introduced in our considerations have an explicit nonrelativistic limit. 
They correspond to the ones generated in the context of the effective field 
theory description~\cite{theory4}.  

The correlation function $\Phi_D $ of Eq.~(\ref{Lagr_D}), modeling the 
distribution of nucleons in the deuteron, has in the full formalism no 
direct connection to the quantum-mechanical wave function of the deuteron.
Comparison of our formalism to the nonrelativistic 
approaches, dealing with the deuteron wave function, can be performed only 
on the level of matrix elements. In this vein the expectation value of the 
two-body operator $J_\mu^{4N; 2}(q)$ between deuteron states incorporates the 
$D$-wave admixture to the deuteron wave function in the context of potential 
models. This admixture is necessary to explain the quadrupole moment of 
the deuteron. In the context of effective field theory~\cite{theory4} 
it was shown before that the nonrelativistic analog  
of the $J_\mu^{4N; 2}(q)$ operator explains the quadrupole moment/form factor 
of the deuteron. Therefore, we introduce this operator in consistency 
with previous observations.

\section{Numerical results} 

To describe the EM form factors of the deuteron we have the following 
input: the size parameter $\Lambda_D$, describing the distribution 
of nucleons in the deuteron, the EM form factors of the nucleons 
$F_{1,2}^{p,n}(q^2)$ and a set of parameters in the two--body operators 
-- couplings constants $g_i$ and form factors $F_i^{NN}(q^2)$. 
In fixing the value of the parameter $\Lambda_D$ we use the following 
constraint. In the non--relativistic approximation the vertex function 
$\tilde\Phi_D(-p^2)$  represents the wave function of the deuteron. 
Therefore, a constraint condition for $\Lambda_D$ is set by the
deuteron size that, according to potential model 
calculations, is bound as $\la | r^{-2} | \ra  < 0.02$ GeV$^2$~\cite{theory1}. 
Employing this condition, we expect that $\Lambda_D$ is less than 0.5~GeV.

For the set of the EM nucleon form factors $F_{1,2}^{p,n}(q^2)$ we use
parametrizations in a fit to the corresponding data. 
We use two available forms -- 
the Mergell-Meissner-Drechsel (MMD) parametrization~\cite{Mergell:1995bf} 
and the Kelly parametrization~\cite{Kelly:2004hm}. We stress that 
when we restrict to the use of one--body transition operators only we cannot 
reproduce the data on the EM form factors of the deuteron. In particular, 
we cannot correctly predict the $Q^2$-dependence of the deuteron form 
factors -- the nodes of the charge and magnetic form factors  
at about 0.7~GeV$^2$ and 2~GeV$^2$~\cite{TomasiGustafsson:2005ni} 
are not reproduced. Also, the quadrupole form factor cannot be explained --
with the one--body operators present only its normalization is completely
underestimated.
As is known from potential models~\cite{theory1} the $D$-wave 
component in the deuteron wave function mainly contributes to the quadrupole 
moment and the quadrupole moment vanishes if only the S-wave deuteron wave 
function is considered. Because at this level our
correlation function $\tilde\Phi_D(-p^2)$ does not contain an explicit 
admixture of a $D$--wave component, the result for the quadrupole moment 
with one--body operators is negligibly small when compared to data.

In the present approach we propose and test if a suitable choice for the
additional two--body operators can reproduce the full quantitative
structure of the deuteron form factors. In particular, as will be shown, the
two--body operator $J_\mu^{4N; 2}(q)$ will give the dominant contribution 
on top of the one--body structures to fully explain the quadrupole form factor.
We proceed by fixing the related parameters $g_2$ and 
$F_2^{NN}(q^2)$ of $J_\mu^{4N; 2}(q)$ 
to reproduce the measured quadrupole form factor of 
the deuteron. The couplings $g_{1,3}$ and the form factors $F_{1,3}^{NN}(Q^2)$
of the additional two--body operators are 
adjusted to obtain a refined fit to the charge and magnetic form 
factors, including the nodes at about 0.7 and 
2~GeV$^2$~\cite{TomasiGustafsson:2005ni}. For the form factors 
$F_{1,2,3}^{NN}(Q^2)$ we use the following parametrization: 
\eq 
F_{1,3}^{NN}(Q^2)=\frac{Q^2}{\Lambda^2+Q^2}
\exp\Big(-\frac{Q^2}{\Lambda^2_1}\Big) \,, \hspace*{2cm} 
F_2^{NN}(Q^2)=\frac{\Lambda^2}{\Lambda^2+Q^2} 
\exp\Big(-\frac{Q^2}{\Lambda^2_2}\Big) \,,
\en 
where $\Lambda$, $\Lambda_1$ and $\Lambda_2$ are the size parameters 
(for $F_1^{NN}(Q^2)$ and $F_3^{NN}(Q^2)$ we use the same ones).  
The form factors $F_1^{NN}(Q^2)$ and $F_3^{NN}(Q^2)$ should vanish 
at zero recoil such that the deuteron charge is not renormalized 
(to preserve the charge conservation), while the form factor $F_2^{NN}(Q^2)$
does not affect current conservation and, therefore, does not vanish 
at zero recoil. In Table I we present the results for the fit parameters 
$g_{1,2,3}$, $\Lambda$ and $\Lambda_{1,2}$ using the data on the deuteron 
EM form factors for two different 
parametrizations~\cite{Mergell:1995bf,Kelly:2004hm} of the nucleon Dirac 
and Pauli form factors. 

Our numerical results for the deuteron charge, magnetic and quadrupole 
form factors as well as for the two form factors, $A(Q^2)$ and $B(Q^2)$,
entering in the differential cross section and for the polarization tensor 
$T_{20}$ are shown in Figs.4--18. 
In particular, in Figs.4--6 we present the results of our approach for the  
$G_C(Q^2)$, $G_Q(Q^2)$ and $G_M(Q^2)$ form factors (total contribution 
of the set of the diagrams in Fig.3). In Figs.7--9 we analyze the separate 
contributions of the graphs of Fig.3 to $G_C(Q^2)$, $G_Q(Q^2)$ and $G_M(Q^2)$. 
In Figs.10--15 we show the separate contributions of the two--body operators 
generating the diagram of Fig.3d on a logarithmic scale for 
$|G_C(Q^2)|$, $|G_Q(Q^2)|$ and $|G_M(Q^2)|$ (Figs.10, 12, 14) 
and on a linear  scale for $G_C(Q^2)$, $G_Q(Q^2)$ and $G_M(Q^2)$ 
(Figs.11, 13, 15). In Figs.16 and 17 we plot our 
results for the form factors $A(Q^2)$ and $B(Q^2)$. 
Finally, in Fig.18 we present the results for the deuteron polarization 
tensor $T_{20}(Q^2)$. We select $\theta_e=70^o$ to make
a comparison consistent with the experimental data. We also perform 
a comparison with other theoretical calculations.  
We present our results for the two sets of parametrizations of the nucleon 
EM form factors -- the MMD~\cite{Mergell:1995bf} 
and the Kelly parametrization~\cite{Kelly:2004hm}. We also compare our 
fit to the parametrization derived in Ref.~\cite{TomasiGustafsson:2005ni}
(denoted as TGA parametrization)
and data~\cite{Abbott:2000ak}-\cite{Platchkov:1989ch}. 
Note that there are several other phenomenological parametrizations
for the deuteron EM form factors in the literature -- 
see Refs.~\cite{Abbott:2000ak,Kobushkin:1994ed,Sick}.
Our main observations are: 
first of all, the one--body diagrams in Figs.3(a)-3(c) correctly reproduce 
the normalizations of charge and magnetic form factors. 
The contribution of the one--body diagrams in Figs.3(b) and 3(c),  
generated by gauging the coupling of the deuteron with nucleons is 
strongly suppressed in all form factors. They become almost constant
at $Q^2 > 0.5$ GeV$^2$. Second, the 
contribution of the one--body diagram of Fig.3(a) is dominant in 
the charge and magnetic form factors up to 1 GeV$^2$. Above 1~GeV$^2$ 
the contribution of the two--body operators $J_\mu^{4N; i}$,  
generating the diagrams of Fig.3(d), becomes important for the charge and 
magnetic form factors. In the case of the quadrupole form factor 
the contribution of the two--body graph  [Fig.3(d)] generated 
by the operator  $J_\mu^{4N; 2}$ is dominant for both the normalization 
and the $Q^2$-dependence of this quantity. The operator $J_\mu^{4N; 2}$ 
simulates the contribution of the $D$--wave component of the deuteron 
wave function. Therefore, inclusion of the two--body operators 
$J_\mu^{4N; i}(q)$ is sufficient 
to describe the deuteron EM form factors as well as elastic $eD$ 
scattering. In particular, the two crossing points of the charge and 
magnetic form factors of the deuteron are also successfully reproduced  
when including the two--body operators. 
Without the two--body operators, one cannot correctly reproduce  
the two nodes at $Q^2\sim$ 0.7 GeV$^2$(for $G_C$) and 2 GeV$^2$ (for $G_M$).

We also tested different forms of the deuteron correlation function 
$\Phi_D$ (monopole, dipole, Gauss, etc.). It is found that the physical 
observables are weakly sensitive to its form, whereas they are dominantly 
controlled by the scale parameter of the correlation function. 
Inclusion of additional terms (e.g. with derivatives) in the Lagrangian, 
describing the bound state structure of the 
deuteron [see Eq.~(\ref{Lagr_D})], does not lead to a considerable improvement
in the description of the deuteron observables; it only introduces
additional parameters. Because we are interested in keeping the number of free 
parameters to a minimum while still obtaining a good fit to the data we 
restrict the coupling of the deuteron to the constitutent nucleons to
the simplest form.

\section{Conclusions}

In this work, we applied a relativistic effective Lagrangian approach 
to study the EM properties of the deuteron considering the deuteron 
as a weakly bound state of a proton and a neutron.
We found that in the present approach
two--body interaction terms are crucial to reproduce
the quadrupole moment of the deuteron and the two crossings
of the charge and magnetic form factors at $\sim 0.7$ 
and $\sim 2$~GeV$^2$, respectively.
The effective two--body operators we include reflect and model the $S/D$ 
mixing in the deuteron wave function and pion exchange contributions. 
With the adjusted parameters, listed in Table 1, we obtain a reasonable  
description of the EM form factors of the deuteron up to $Q^2\sim 2$ GeV$^2$ 
using MMD parametrization of the nucleon EM form factors. Note that our result 
for the deuteron polarization tensor $T_{20}$ using the MMD parametrization 
deviates from the data and predictions of theoretical 
approaches~\cite{theory11,theory33,theory34} in the region 
$\simeq 1 - 1.4$ GeV$^2$. 

Finally, recent experiments of the electron--proton polarization 
transfer scattering~\cite{PEPTS} show that two-photon exchange may play 
a role for determing the charge form factor of the proton~\cite{TPE}. It is
therefore of great 
interest to check if the two--photon exchange mechanism also plays
a prominent role on the form factors of 
the deuteron~\cite{Dong:2006wm} as well as of the neutron. Work along this
line is currently in progress.        

\begin{acknowledgments}

This work was supported by the DFG under Contract Nos. FA67/31-1 and
GRK683, National Sciences Foundations Grant No. 10775148 and 
CAS Grant No. KJCX3-SYW-N2 (Y.B.D.). This research is also part of 
the EU Integrated Infrastructure Initiative Hadronphysics project under  
Contract No. RII3-CT-2004-506078 and the President Grant of Russia 
``Scientific Schools''  No. 871.2008.2. Y.B.D. thanks the 
T\"ubingen theory group for its hospitality.  

\end{acknowledgments}

\vspace*{2cm} 

\begin{center}
{\bf Table 1.} Parameters defining the two--body EM nucleon operators 
extracted from data using two forms of the parametrization of 
the Dirac and Pauli nucleon form factors: MMD~\cite{Mergell:1995bf} 
and Kelly~\cite{Kelly:2004hm}. 

\vspace*{0.5cm}
\begin{tabular} {|l|l|l|l|l|l|l|l|} 
\hline   
$F^{p,n}_{1,2}(Q^2)$ 
&$\Lambda_D$ (GeV) 
&$g_1$ (fm$^4$) 
&$g_2$ (fm$^5$) 
&$g_3$ (fm$^4$) 
&$\Lambda$ (GeV)  
&$\Lambda_{1}$ (GeV)  
&$\Lambda_{2}$ (GeV)  \\ 
\hline
MMD~\cite{Mergell:1995bf} 
&0.25 & 0.06 & -1.97 & 0.16 & 1 & 0.70 & 0.80\\ \hline
Kelly~\cite{Kelly:2004hm} 
&0.30 & 0.23 & -1.48 & 0.06 & 1 & 0.58 & 0.58\\ \hline
\end{tabular}
\end{center}

\newpage 

\begin{figure}
\begin{center}
\epsfig{file=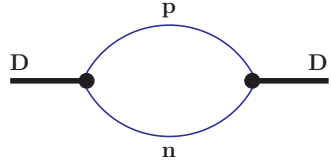, scale=.6}
\caption{\footnotesize (Color online). Deuteron mass operator} 
\end{center} 
\end{figure}

\begin{figure}
\vspace*{1.5cm} 
\begin{center}
\epsfig{file=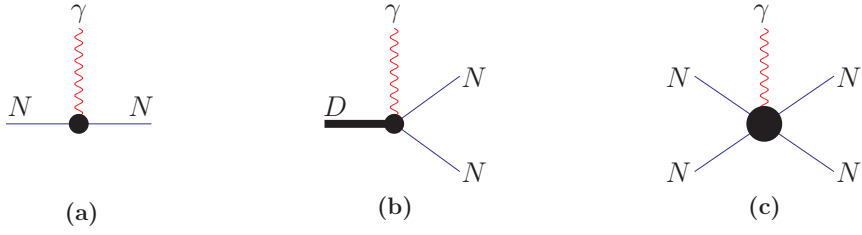, scale=.75}
\caption{\footnotesize (Color online). Electromagnetic operators} 
\end{center} 
\end{figure}

\begin{figure} 
\vspace*{.5cm} 
\begin{center}
\epsfig{file=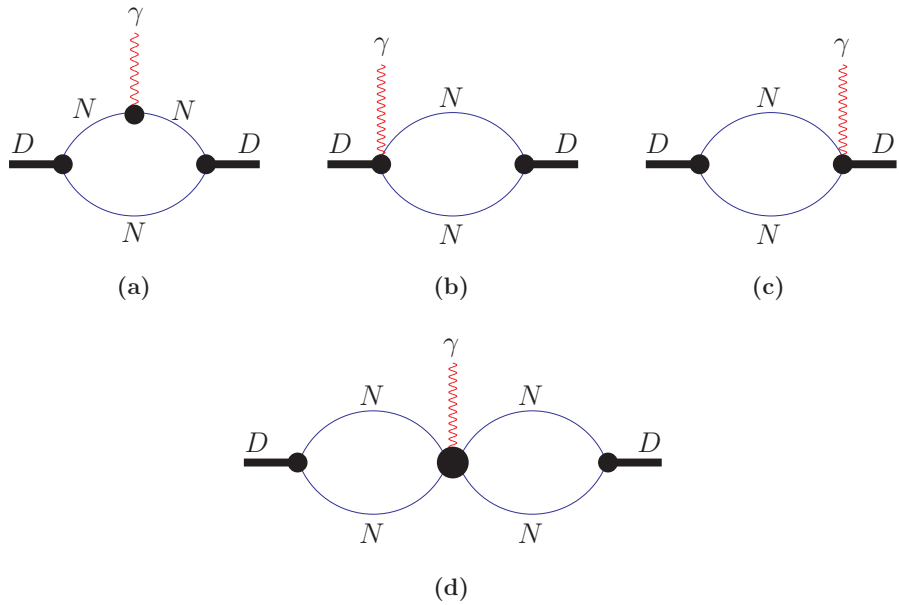,scale=.75}
\caption{\footnotesize (Color online). 
One-- and two--body diagrams contributing to 
the electromagnetic deuteron form factors.} 
\end{center} 
\end{figure}

\newpage 
\begin{figure} 
\vspace*{1.25cm} 
\centering 
\includegraphics[scale=1]{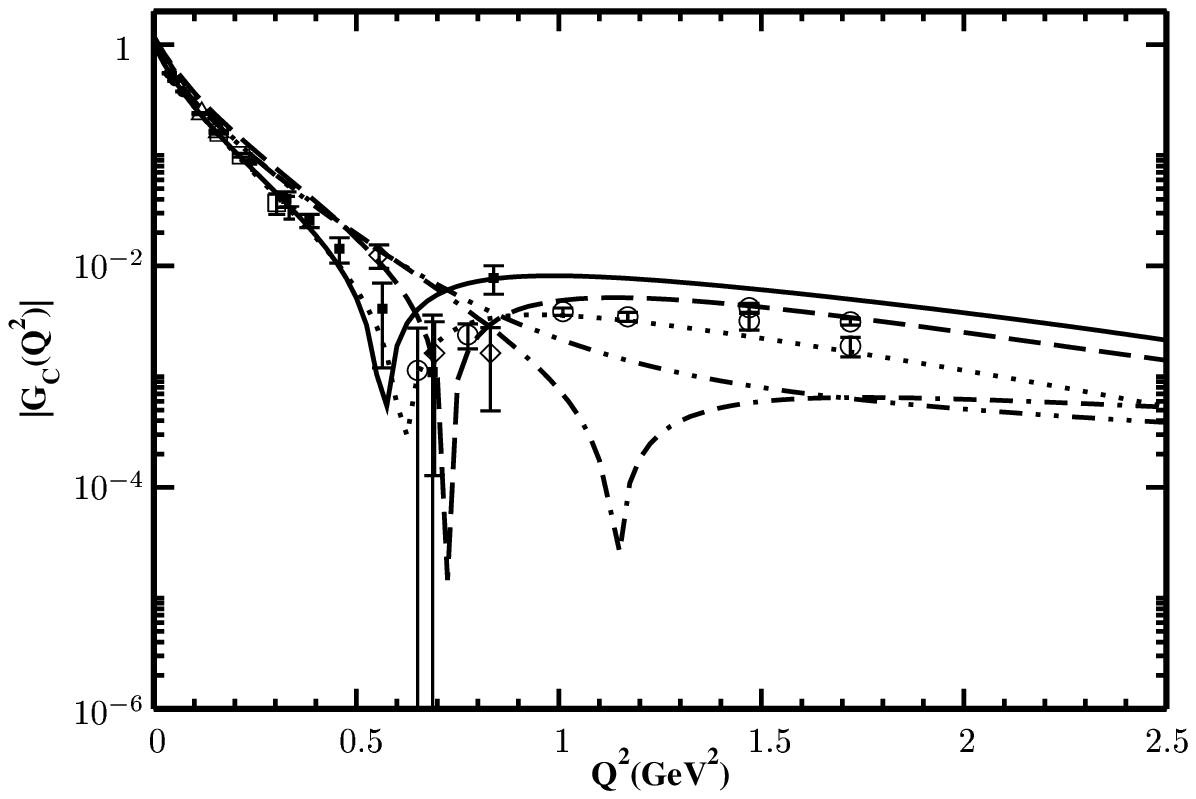}
\caption{\footnotesize 
Form factor $\mid G_C(Q^2)\mid $. 
The solid curve is the result of the TGA parametrization. 
The double dash--dotted and dashed lines are our
results with the MMD~\cite{Mergell:1995bf} parametrization  
restricting to one--body and including two--body electromagnetic currents, 
respectively. The double dot-dashed and dotted lines are our results 
with the Kelly~\cite{Kelly:2004hm} parametrization  
restricting to one--body and including two--body electromagnetic currents, 
respectively. The data are from~\cite {Abbott:2000fg} (open circle), 
\cite{Bouwhuis:1998jj}  (open square), \cite{Garcon:1993vm} (open diamond), 
\cite{Gilman:1990vg} (Plux), \cite{Schulze:1984ms} 
(triangle up), \cite{Voitsekhovsky:1986xh} (filled circle), and 
\cite{Nikolenko:2003zq} (filled square). }
\end{figure}

\begin{figure} 
\vspace*{1cm}
\centering
\includegraphics[scale=1]{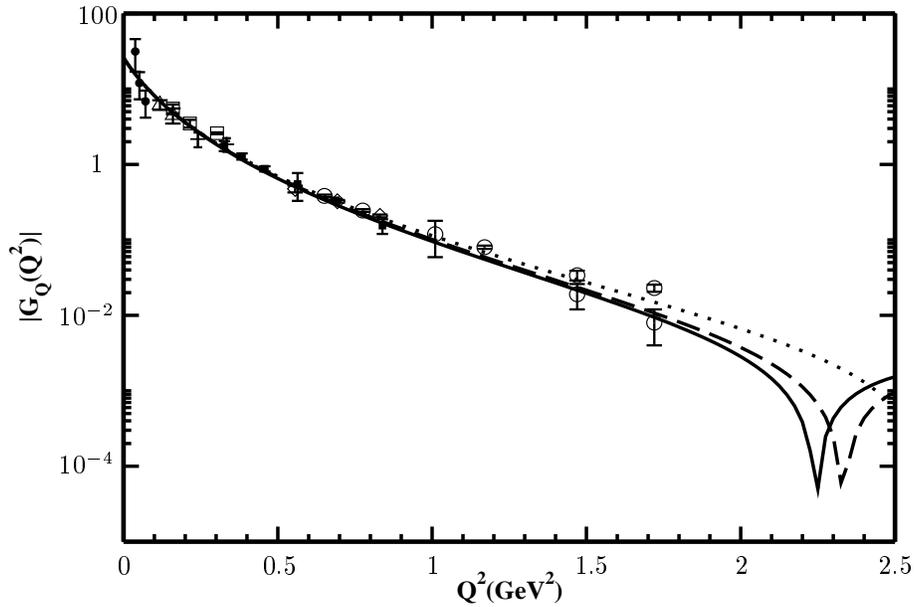}
\caption{\footnotesize 
Form factor $\mid G_Q(Q^2)\mid$. 
Notations are the same as in Fig.4.}
\end{figure}

\newpage 

\begin{figure} 
\centering
\vspace*{1.1cm} 
\includegraphics[scale=1]{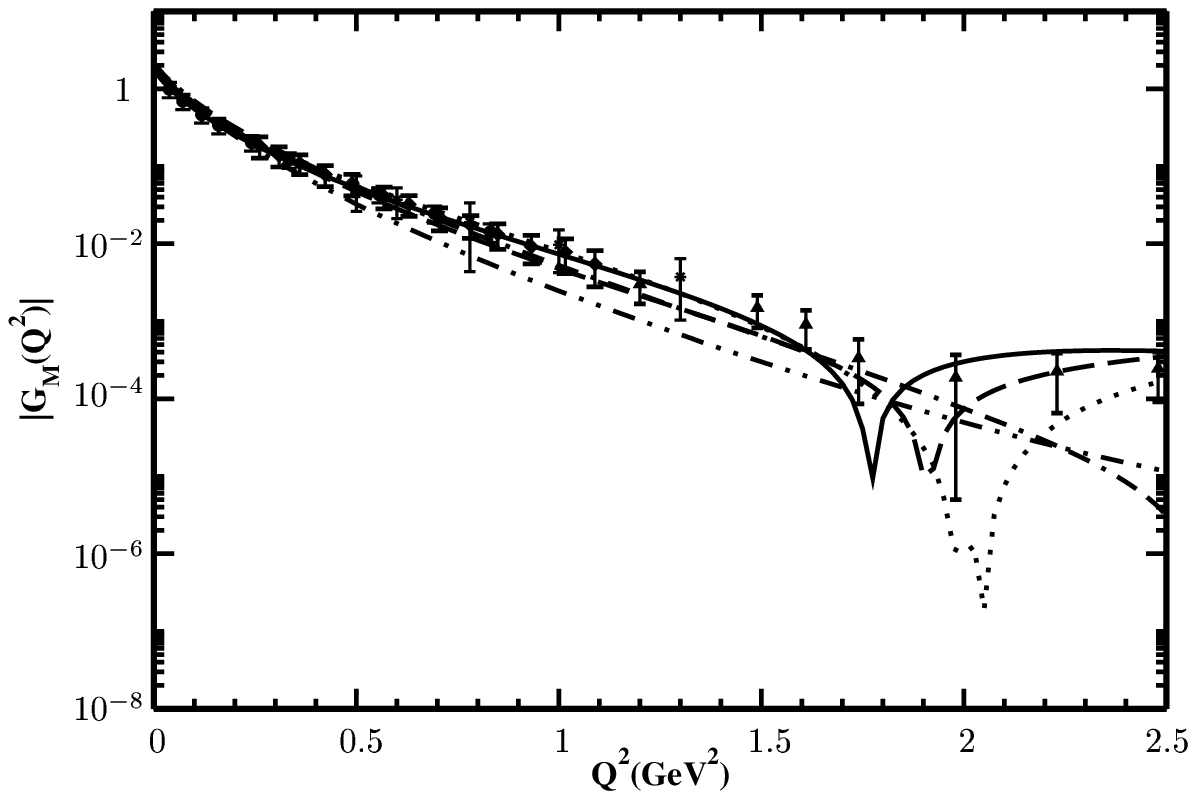}
\caption{\footnotesize 
Form factor $\mid G_M(Q^2)\mid$. 
The solid curve is the result of the TGA parametrization. 
The double dash--dotted and dashed lines are our
results with the MMD~\cite{Mergell:1995bf} parametrization 
restricting to one--body and including two--body electromagnetic currents, 
respectively. The double dot-dashed and dotted lines are our results 
with the Kelly~\cite{Kelly:2004hm} parametrization 
restricting to one--body and including two--body electromagnetic currents, 
respectively. The data are 
quoted from~\cite{Garcon:1993vm}(circle), \cite{Auffret:1985tg} (diamond), 
\cite{Bosted:1989hy} (square), \cite{Cramer:1986kv} (star).}
\end{figure}

\begin{figure} 
\centering
\vspace*{1cm}
\includegraphics[scale=1]{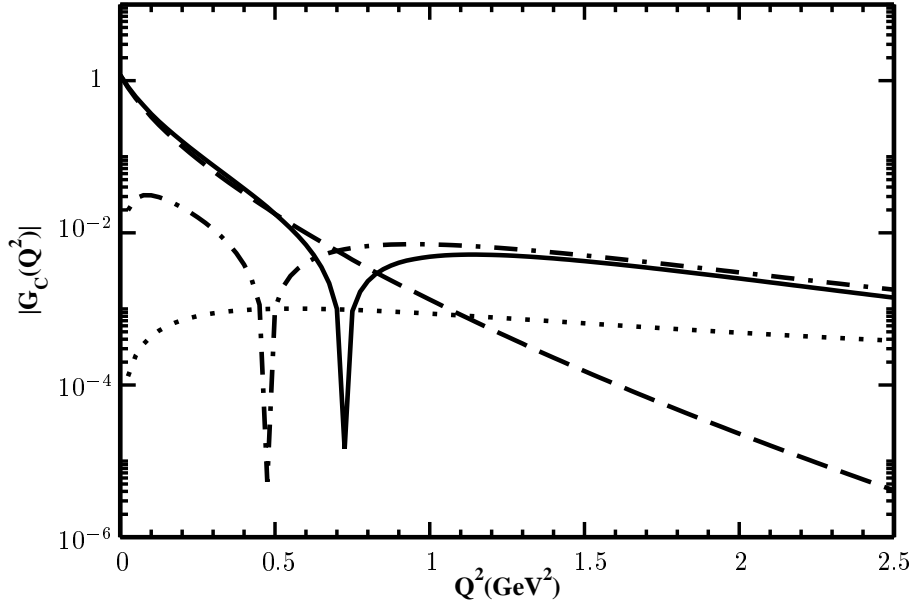}
\caption{\footnotesize 
Form factor $\mid G_C(Q^2)\mid$. Separate 
contributions of the diagrams in Fig.3 with the 
MMD~\cite{Mergell:1995bf} parametrization: diagram 
in Fig.3a (the dashed line), diagrams in Figs.3b and 3c (the dotted line), 
the total contribution of two--body operators generating the 
diagram Fig.3d (the dot--dashed line).} 
\end{figure}

\newpage 

\begin{figure}
\vspace*{1.25cm} 
\centering
\includegraphics[scale=1]{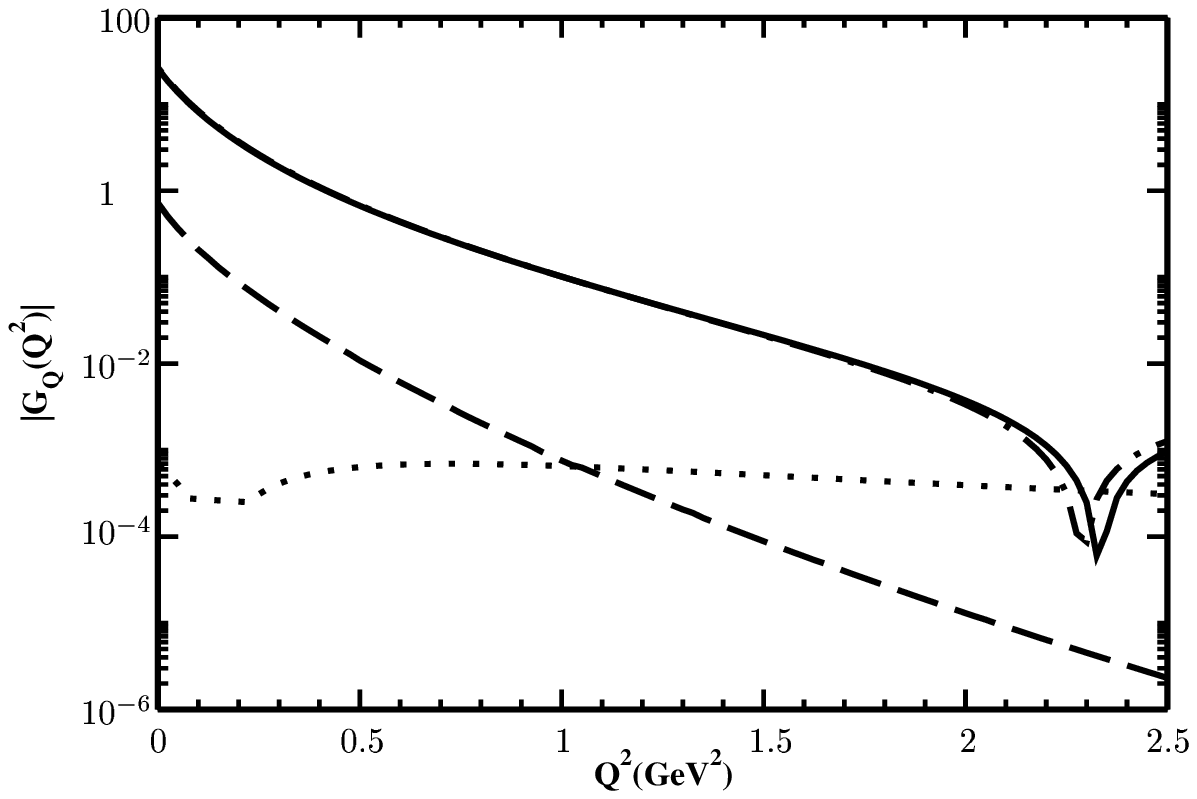}
\caption{\footnotesize 
Form factor $\mid G_Q(Q^2)\mid$. Separate contributions 
of the diagrams in Fig.3 with the MMD~\cite{Mergell:1995bf} 
parametrization. \\ 
Notations are the same as in Fig.7.} 
\end{figure}

 \begin{figure}
\vspace*{1cm} 
\centering
\includegraphics[scale=1]{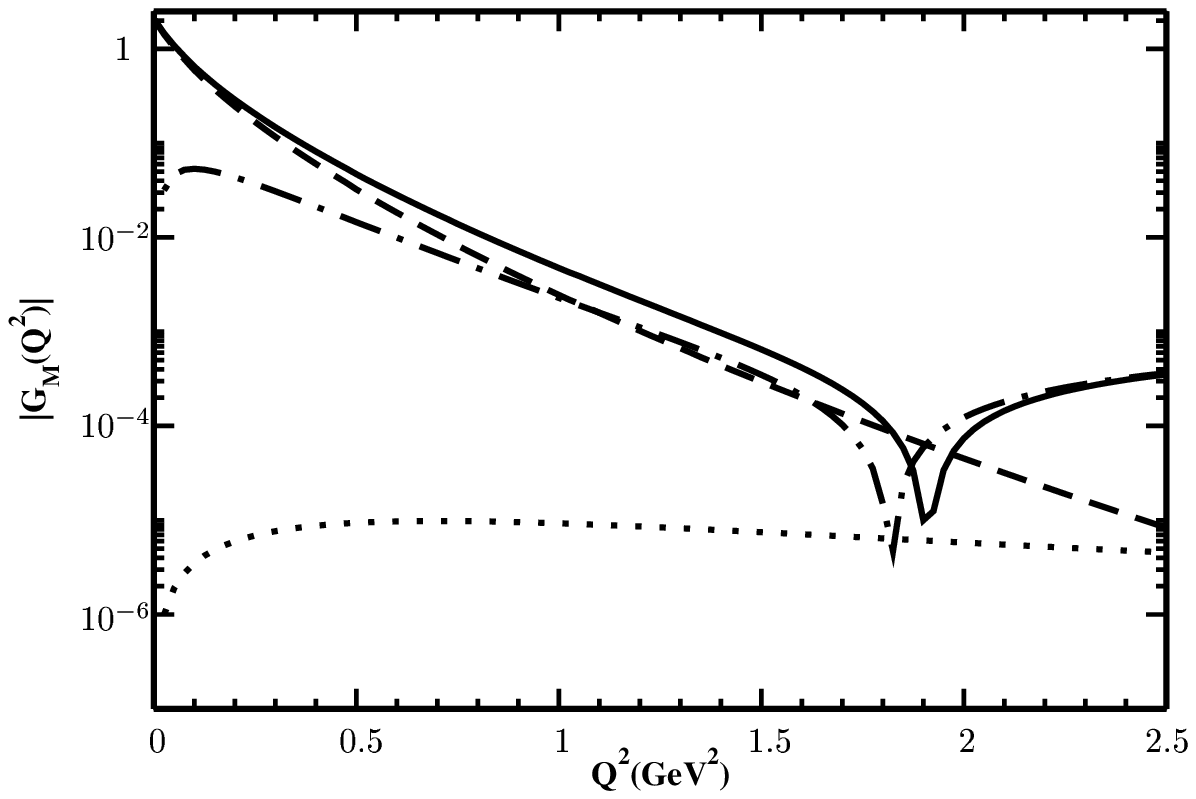}
\caption{\footnotesize 
Form factor $\mid G_M(Q^2)\mid$. Separate contributions 
of the diagrams in Fig.3 with the MMD~\cite{Mergell:1995bf} 
parametrization.\\ 
Notations are the same as in Fig.7.} 
\end{figure}

\newpage 

\begin{figure}
\vspace*{1.25cm} 
\centering
\includegraphics[scale=1]{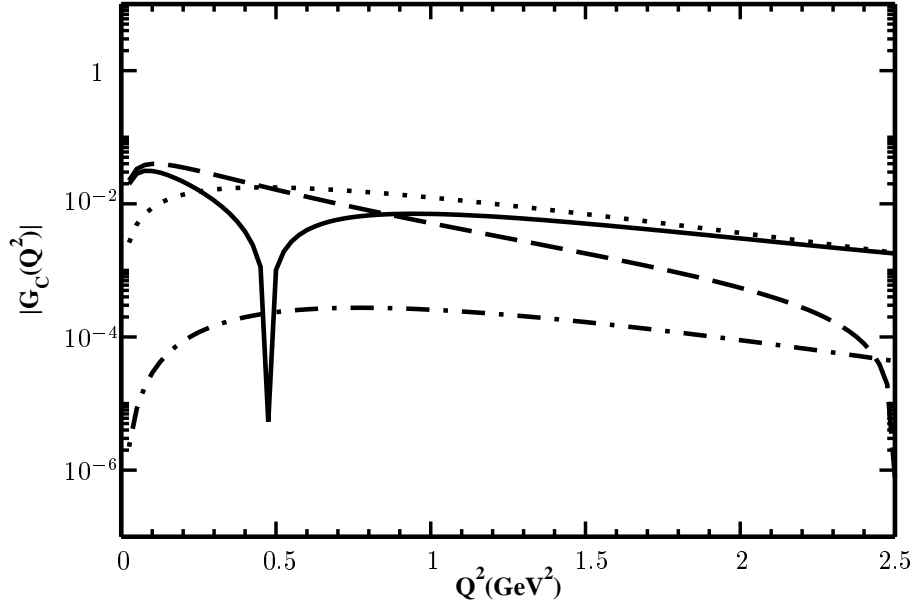} 
\caption{\footnotesize 
Form factor $\mid G_C(Q^2)\mid$. Separate contributions 
of the two--body operators with the MMD~\cite{Mergell:1995bf} 
parametrization: 
the contribution of the $J_\mu^{4N; 1}$ operator (the dotted line), 
the contribution of the $J_\mu^{4N; 2}$ operator (the dashed line), 
the contribution of the $J_\mu^{4N; 3}$ operator (the dot-dashed line)
and the total result (solid line).} 
\end{figure}

\begin{figure}
\vspace*{1cm} 
\centering
\includegraphics[scale=1]{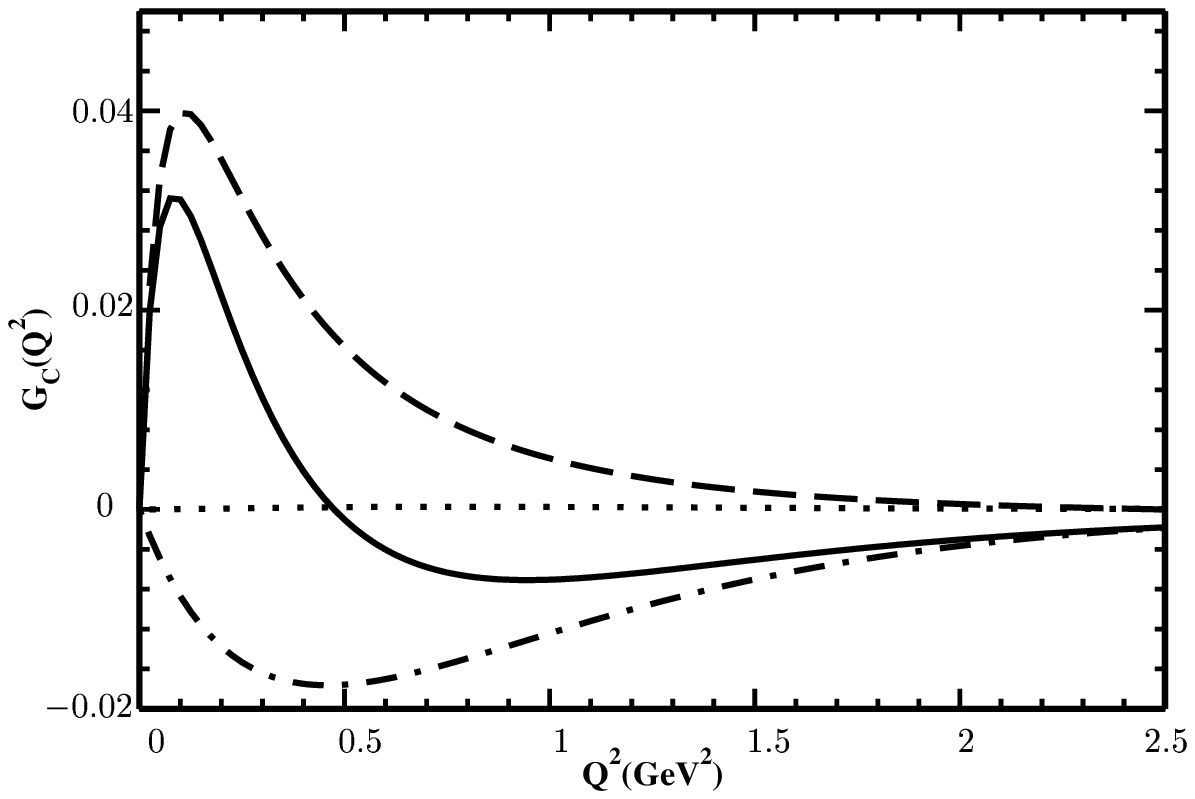}
\caption{\footnotesize 
Form factor $G_C(Q^2)$ on a linear scale. 
Separate contributions of the two--body operators with 
the MMD~\cite{Mergell:1995bf} 
parametrization. Notations are the same as in Fig.10.}
\end{figure}

\newpage 

\begin{figure}
\vspace*{1.25cm} 
\centering
\includegraphics[scale=1]{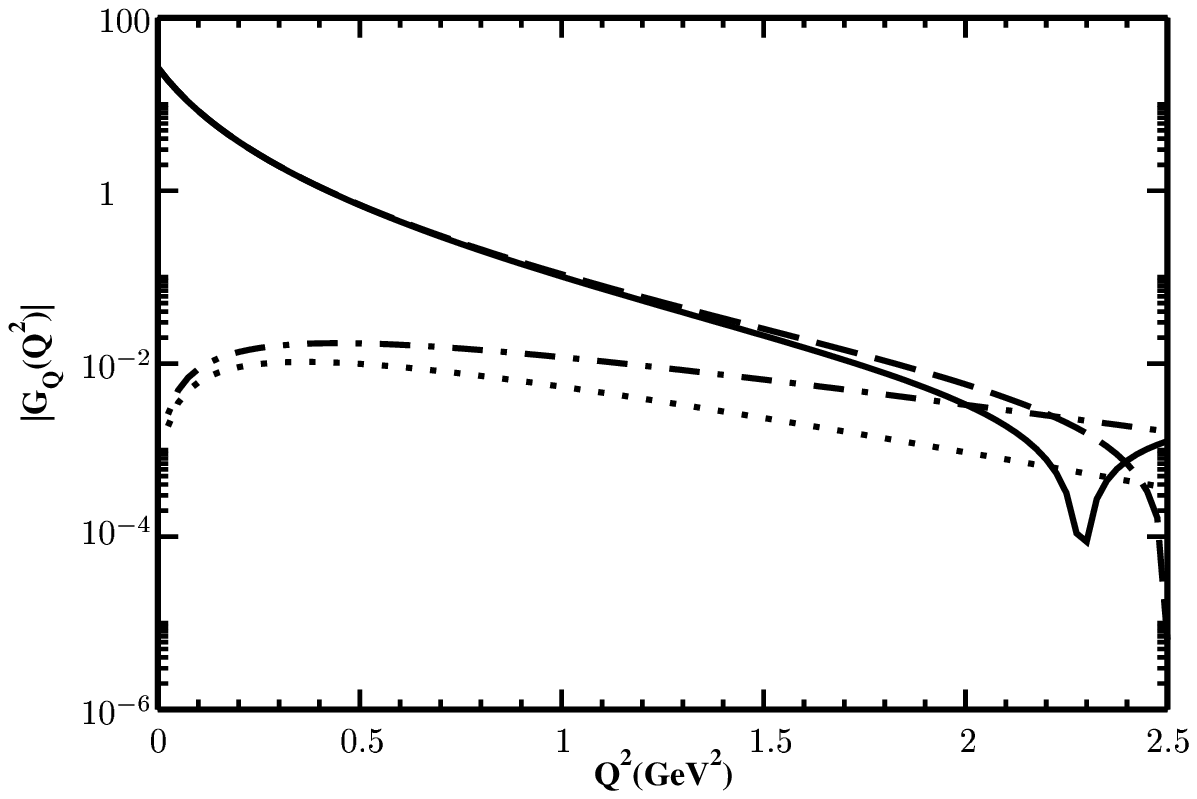}
\caption{\footnotesize 
Form factor $\mid G_Q(Q^2)\mid$. 
Separate contributions of the two--body operators with 
the MMD~\cite{Mergell:1995bf} 
parametrization. Notations are the same as in Fig.10.}
\end{figure}

\begin{figure}
\vspace*{1cm} 
\centering
\includegraphics[scale=1]{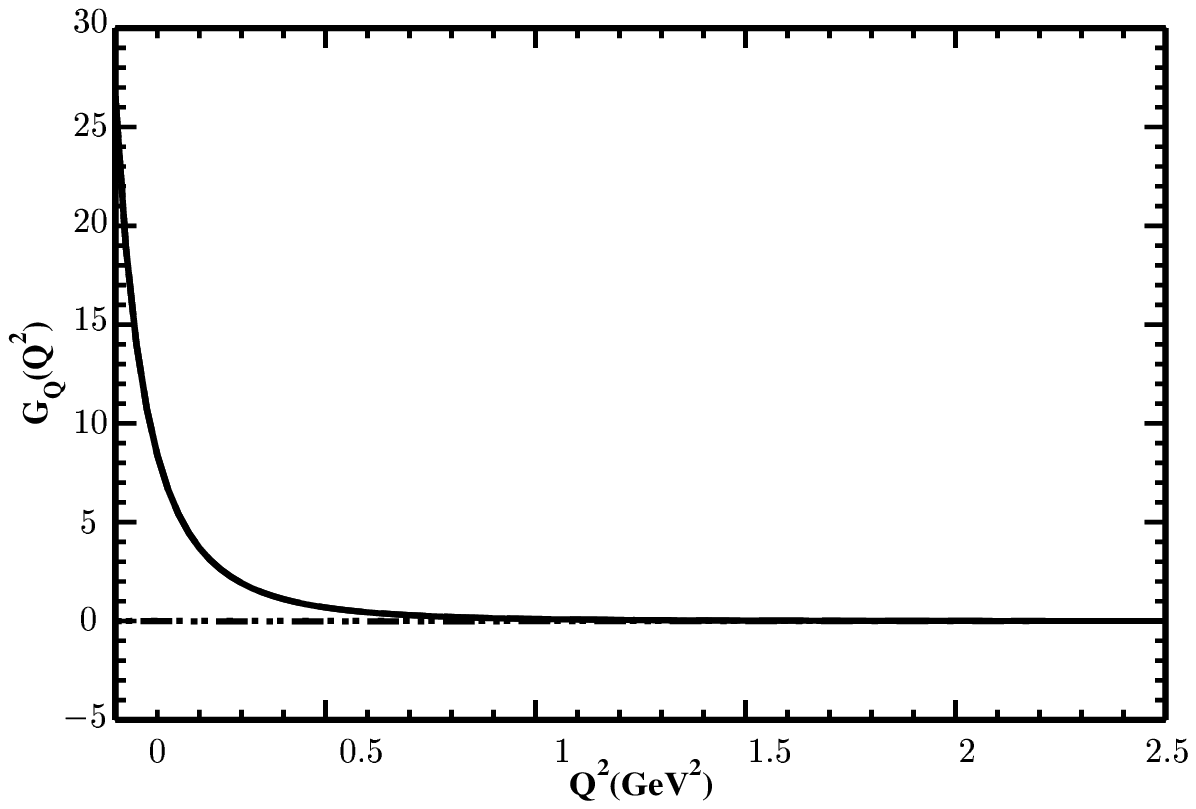}
\caption{\footnotesize 
Form factor $G_Q(Q^2)$ in a linear scale. 
Separate contributions of the two--body operators with 
the MMD~\cite{Mergell:1995bf} 
parametrization. Notations are the same as in Fig.10.}
\end{figure}

\newpage 
\begin{figure}
\vspace*{1.25cm} 
\centering
\includegraphics[scale=1]{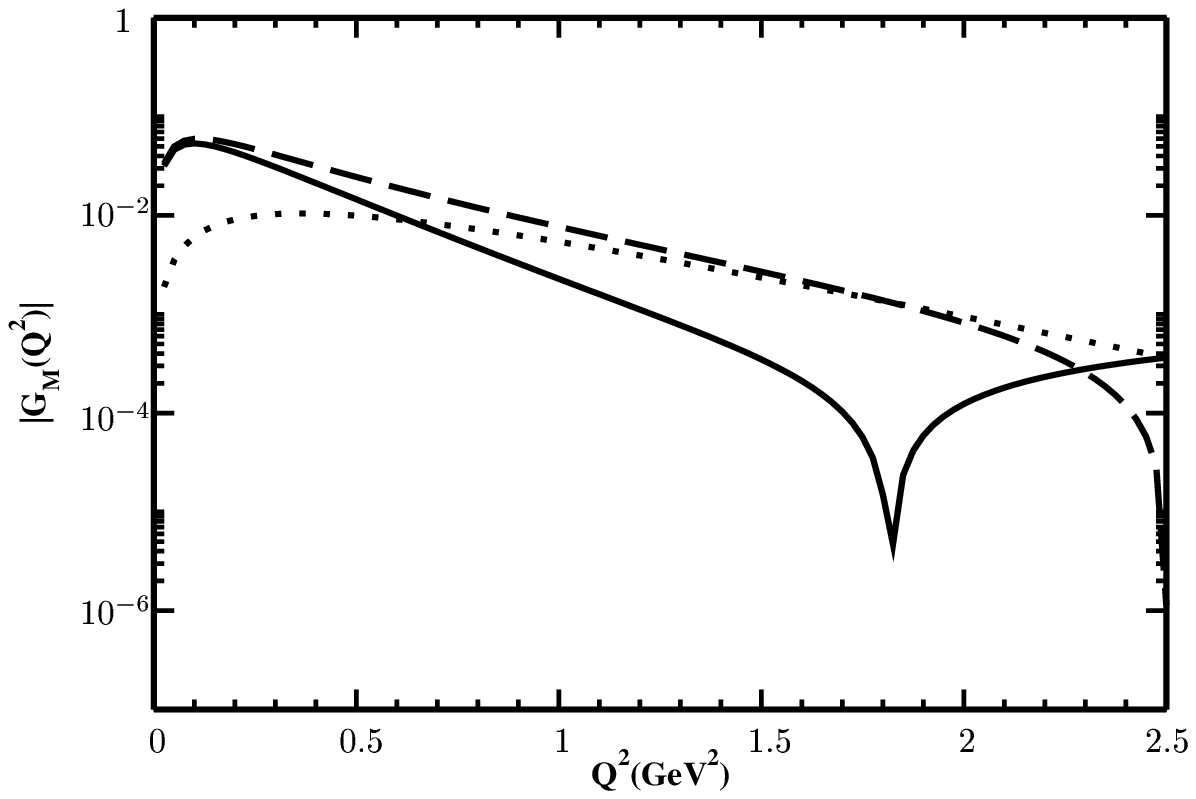}
\caption{\footnotesize 
Form factor $\mid G_M(Q^2)\mid$. 
Separate contributions of the two--body operators with 
the MMD~\cite{Mergell:1995bf} parametrization. 
Notations are the same as in Fig.10.}
\end{figure}

\begin{figure}
\vspace*{1cm} 
\centering
\includegraphics[scale=1]{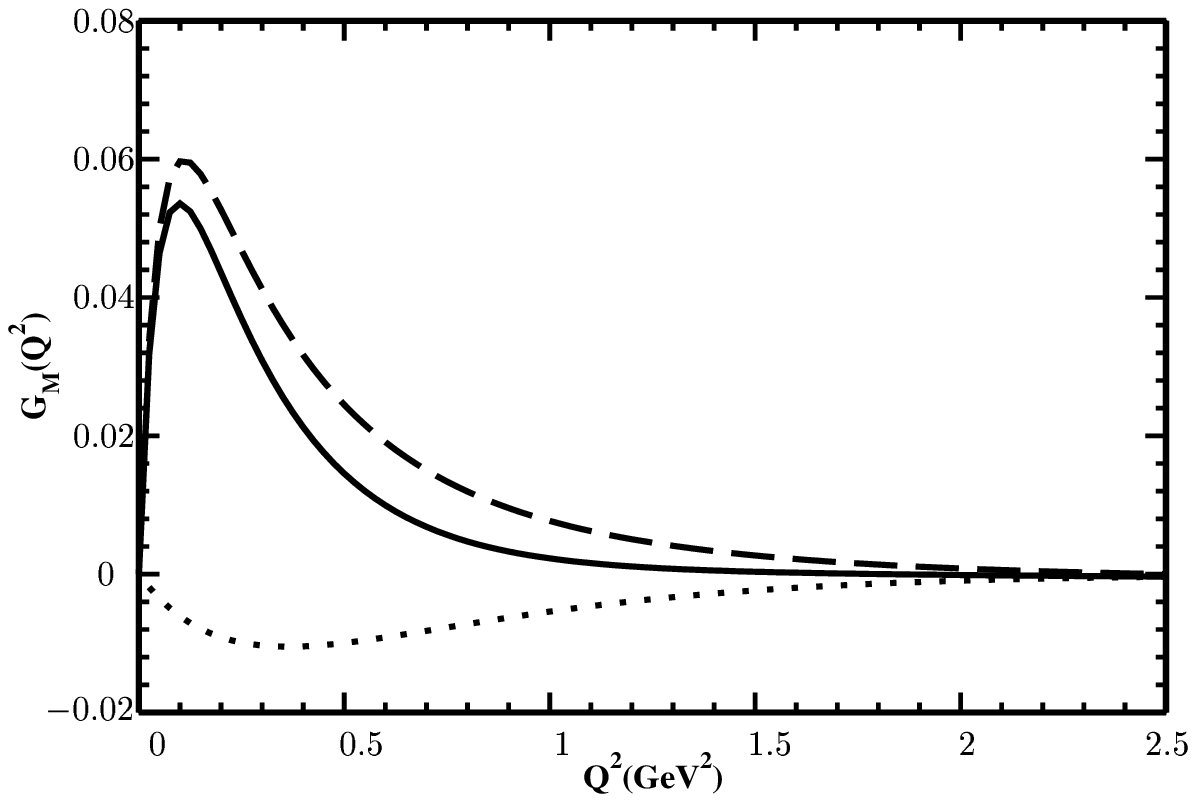}
\caption{\footnotesize 
Form factor $G_M(Q^2)$ on a linear scale. 
Separate contributions of the two--body operators with 
the MMD~\cite{Mergell:1995bf} parametrization. 
Notations are the same as in Fig.10.}
\end{figure}

\newpage 
\begin{figure}
\vspace*{1.25cm} 
\centering
\includegraphics[scale=1]{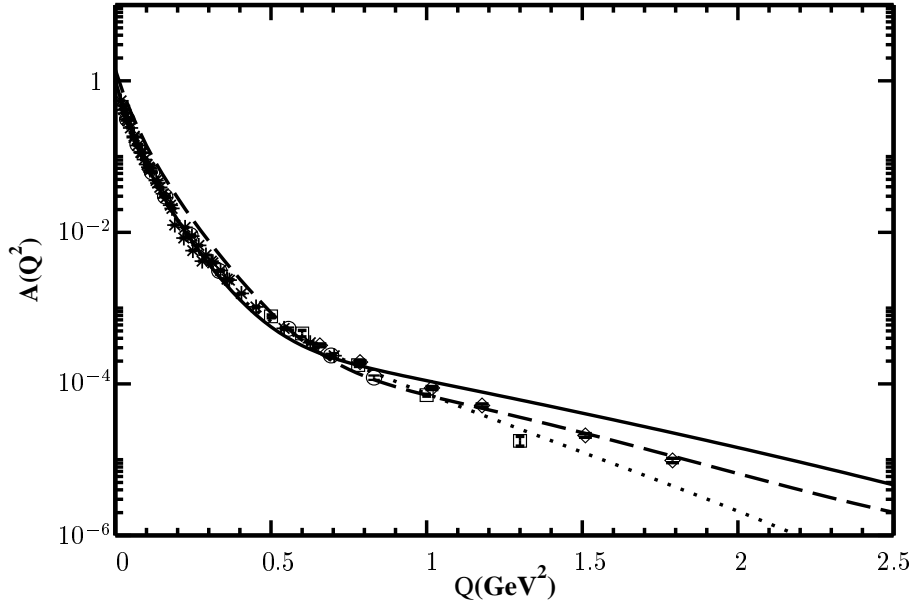}
\caption{\footnotesize 
Form factor $A(Q^2)$. The solid curve is the result of the 
TGA parametrization. The dashed and dotted lines are our
results with the MMD~\cite{Mergell:1995bf} and Kelly~\cite{Kelly:2004hm} 
parametrizations, respectively. 
The data are quoted from~\cite{Garcon:1993vm} (circle), 
\cite{Abbott:1998sp} (diamond), \cite{Cramer:1986kv} (square), 
\cite{Platchkov:1989ch} (star), respectively.}
\end{figure}

\begin{figure} 
\vspace*{1cm} 
\centering
\includegraphics[scale=1]{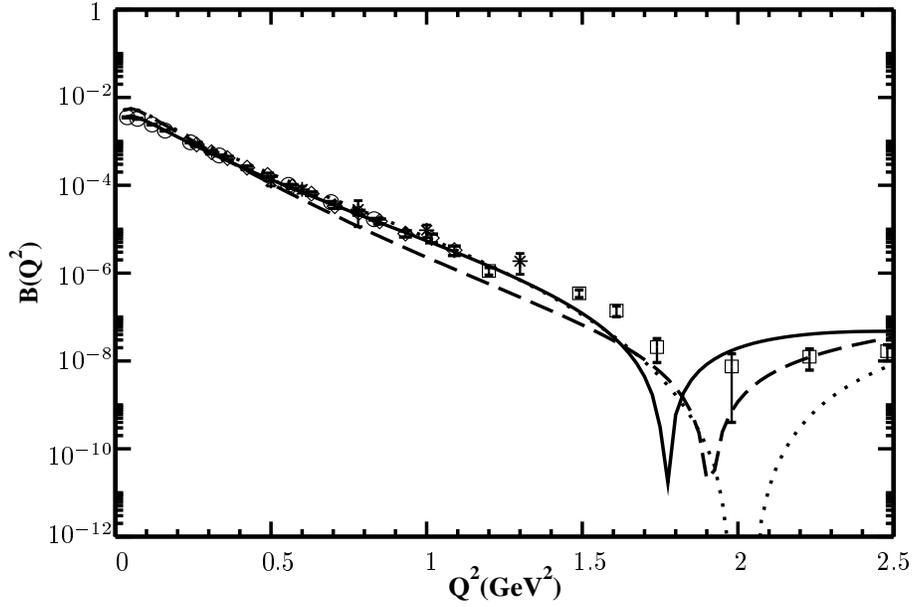}
\caption{\footnotesize 
Form factor $B(Q^2)$. Notations are same as in Fig.16}
\end{figure}

\newpage

\begin{figure} 
\vspace*{1cm} 
\centering
\includegraphics[scale=1]{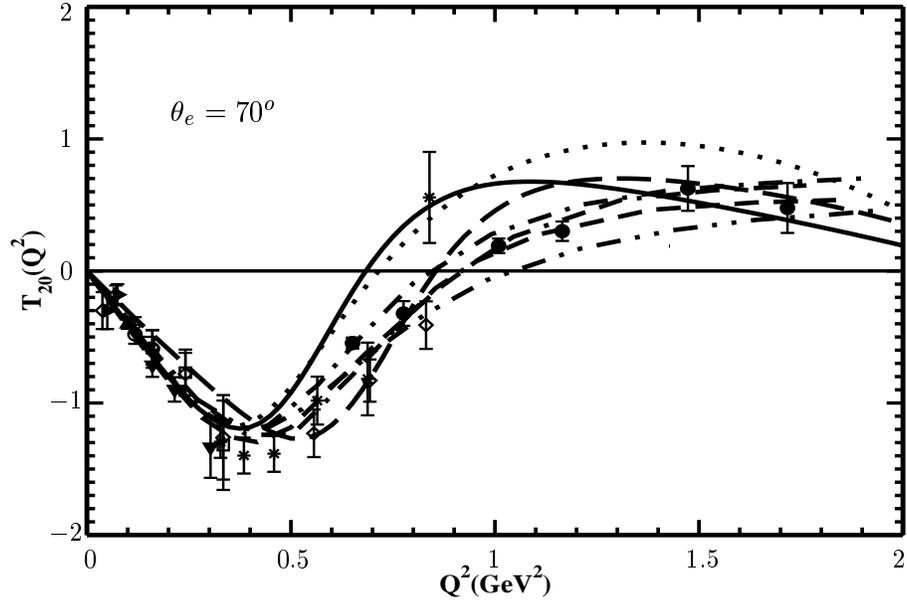}
\caption{\footnotesize 
Deuteron polarization tensor $T_{20}(Q^2)$ at $\theta_e = 70^o$. 
The solid curve is the result of the TGA parametrization. 
The dashed and dotted lines are our
results with the MMD~\cite{Mergell:1995bf} and 
Kelly~\cite{Kelly:2004hm} parametrizations, respectively. 
The data are quoted from~\cite{Schulze:1984ms} (circle),  
\cite{Dmitriev:1985us} (triangle right),  
\cite{Gilman:1990vg} (square),  
\cite{Garcon:1993vm} (diamond),  
\cite{FerroLuzzi:1996dg} (filled triangle up),  
\cite{Bouwhuis:1998jj} (filled triangle down),  
\cite{Abbott:2000fg} (filled circle), 
\cite{Nikolenko:2003zq} (star). 
The theoretical results are taken from~\cite{theory11} 
(dot--dashed line), \cite{theory33} (short--dashed line), 
\cite{theory34} (double dash--dotted line), 
\cite{theory12} (double dotted--dashed line).} 
\end{figure}


\newpage 

\begin{thebibliography}{99} 

\bibitem{Garcon:2001sz}
  M.~Garcon and J.~W.~Van Orden,
  Adv.\ Nucl.\ Phys.\  {\bf 26}, 293 (2001). 
\bibitem{Gilman:2001yh}
  R.~A.~Gilman and F.~Gross,
  J.\ Phys.\ G {\bf 28}, R37 (2002). 
\bibitem{Sick:2001rh}
  I.~Sick,
  Prog.\ Part.\ Nucl.\ Phys.\  {\bf 47}, 245 (2001). 
\bibitem{Gross:2002ge}
  F.~Gross,
  Eur.\ Phys.\ J.\  A {\bf 17}, 407 (2003). 
\bibitem{Rosenbluth:1950yq}
  M.~N.~Rosenbluth,
  Phys.\ Rev.\  {\bf 79}, 615 (1950).
\bibitem{Arnold:1979cg}
  R.~G.~Arnold, C.~E.~Carlson and F.~Gross,
  Phys.\ Rev.\  C {\bf 21}, 1426 (1980); 
  Phys.\ Rev.\  C {\bf 23}, 363 (1981).
\bibitem{theory1} 
  J.~F.~Mathiot,
  Phys.\ Rep.\  {\bf 173}, 63 (1989); 
  H.~Henning, J.~J.~Adam, P.~U.~Sauer and A.~Stadler,
  Phys.\ Rev.\  C {\bf 52}, R471 (1995); 
  J.~J.~Adam and H.~Arenhovel,
  Nucl.\ Phys.\  A {\bf 614}, 289 (1997). 
\bibitem{theory11} 
  R.~B.~Wiringa, V.~G.~J.~Stoks and R.~Schiavilla,
  Phys.\ Rev.\  C {\bf 51}, 38 (1995). 
\bibitem{theory12} 
  H.~Arenhovel, F.~Ritz and T.~Wilbois,
  Phys.\ Rev.\  C {\bf 61}, 034002 (2000).
\bibitem{theory2}
  M.~Gari and H.~Hyuga,
  Nucl.\ Phys.\  A {\bf 278}, 372 (1977); 
  V.~V.~Burov, S.~M.~Dorkin and V.~N.~Dostovalov,
  Z.\ Phys.\  A {\bf 315}, 205 (1984); 
  V.~V.~Burov and V.~N.~Dostovalov,
  Z.\ Phys.\  A {\bf 326}, 245 (1987); 
  A.~Buchmann, Y.~Yamauchi and A.~Faessler,
  Nucl.\ Phys.\  A {\bf 496}, 621 (1989); 
  H.~Ito and L.~S.~Kisslinger,
  Phys.\ Rev.\  C {\bf 40}, 887 (1989); 
  E.~Hummel and J.~A.~Tjon,
  Phys.\ Rev.\ Lett.\  {\bf 63}, 1788 (1989); 
  G.~Ramalho, M.~T.~Pena and F.~Gross,
  Eur.\ Phys.\ J.\  A {\bf 36}, 329 (2008). 
\bibitem{theory3} 
  V.~A.~Karmanov and A.~V.~Smirnov,
  Nucl.\ Phys.\  A {\bf 575}, 520 (1994); 
  J.~W.~Van Orden, N.~Devine and F.~Gross,
  Phys.\ Rev.\ Lett.\  {\bf 75}, 4369 (1995); 
  J.~Carbonell, B.~Desplanques, V.~A.~Karmanov and J.~F.~Mathiot,
  Phys.\ Rep.\  {\bf 300}, 215 (1998); 
  T.~W.~Allen, W.~H.~Klink and W.~N.~Polyzou,
  Phys.\ Rev.\  C {\bf 63}, 034002 (2001); 
  T.~W.~Allen, G.~L.~Payne and W.~N.~Polyzou,
  Phys.\ Rev.\  C {\bf 62}, 054002 (2000); 
  F.~M.~Lev, E.~Pace and G.~Salme,
  Phys.\ Rev.\  C {\bf 62}, 064004 (2000). 
\bibitem{theory33}
J.~Carbonell and V.~A.~Karmanov,
  Eur.\ Phys.\ J.\  A {\bf 6}, 9 (1999). 
\bibitem{theory34}
  D.~R.~Phillips, S.~J.~Wallace and N.~K.~Devine,
  Phys.\ Rev.\  C {\bf 58}, 2261 (1998). 
\bibitem{theory4}
  D.~B.~Kaplan, M.~J.~Savage and M.~B.~Wise,
  Phys.\ Rev.\  C {\bf 59}, 617 (1999); 
  T.~S.~Park, K.~Kubodera, D.~P.~Min and M.~Rho,
  Phys.\ Rev.\  C {\bf 58}, R637 (1998); 
  J.~W.~Chen, H.~W.~Griesshammer, M.~J.~Savage and R.~P.~Springer,
  Nucl.\ Phys.\  A {\bf 644}, 245 (1998); 
  M.~Walzl and U.~G.~Meissner,
  Phys.\ Lett.\  B {\bf 513}, 37 (2001); 
  D.~R.~Phillips,
  Phys.\ Lett.\  B {\bf 567}, 12 (2003); 
  S.~R.~Beane, M.~Malheiro, J.~A.~McGovern, D.~R.~Phillips and U.~van Kolck,
  Nucl.\ Phys.\  A {\bf 747}, 311 (2005); 
  D.~Choudhury and D.~R.~Phillips,
  Phys.\ Rev.\  C {\bf 71}, 044002 (2005); 
  R.~P.~Hildebrandt, H.~W.~Griesshammer, T.~R.~Hemmert and D.~R.~Phillips,
  Nucl.\ Phys.\  A {\bf 748}, 573 (2005); 
  D.~R.~Phillips,
  J.\ Phys.\ G {\bf 34}, 365 (2007); 
  M.~P.~Valderrama, A.~Nogga, E.~Ruiz Arriola and D.~R.~Phillips,
  Eur.\ Phys.\ J.\  A {\bf 36}, 315 (2008). 
\bibitem{Ivanov:1995zp}
  A.~N.~Ivanov, N.~I.~Troitskaya, M.~Faber and H.~Oberhummer,
  Phys.\ Lett.\  B {\bf 361}, 74 (1995). 
\bibitem{Weinberg:1962hj}
  S.~Weinberg,
  Phys.\ Rev.\  {\bf 130}, 776 (1963);
  A.~Salam,
  Nuovo Cimento\  {\bf 25}, 224 (1962);
  K.~Hayashi, M.~Hirayama, T.~Muta, N.~Seto and T.~Shirafuji,
  Fortsch.\ Phys.\ {\bf 15}, 625 (1967).
\bibitem{Efimov:1993ei}
G.~V.~Efimov and M.~A.~Ivanov,
{\it The Quark Confinement Model of Hadrons},
(IOP Publishing, Bristol $\&$ Philadelphia, 1993). 
\bibitem{Ivanov:1996pz}  
  M.~A.~Ivanov, M.~P.~Locher and V.~E.~Lyubovitskij,
  Few Body Syst.\  {\bf 21}, 131 (1996); 
  M.~A.~Ivanov, V.~E.~Lyubovitskij, J.~G.~K\"orner and P.~Kroll,
  Phys.\ Rev.\ D {\bf 56}, 348 (1997); 
  M.~A.~Ivanov, J.~G.~K\"orner and V.~E.~Lyubovitskij,
  Phys.\ Lett.\ B {\bf 448}, 143 (1999); 
  M.~A.~Ivanov, J.~G.~K\"orner, V.~E.~Lyubovitskij and A.~G.~Rusetsky,
  Phys.\ Rev.\ D {\bf 60}, 094002 (1999); 
  A.~Faessler, T.~Gutsche, M.~A.~Ivanov, J.~G.~K\"orner
  and V.~E.~Lyubovitskij,
  Phys.\ Lett.\ B {\bf 518}, 55 (2001); 
  A.~Faessler, T.~Gutsche, M.~A.~Ivanov, J.~G.~Korner,
  V.~E.~Lyubovitskij, D.~Nicmorus and K.~Pumsa-ard,
  Phys.\ Rev.\ D {\bf 73}, 094013 (2006); 
  A.~Faessler, T.~Gutsche, B.~R.~Holstein, V.~E.~Lyubovitskij,
  D.~Nicmorus and K.~Pumsa-ard,
  Phys.\ Rev.\ D {\bf 74}, 074010 (2006). 
  \bibitem{Faessler:2007gv}
  A.~Faessler, T.~Gutsche, V.~E.~Lyubovitskij and Y.~L.~Ma,
  Phys.\ Rev.\  D {\bf 76}, 014005 (2007); 
  A.~Faessler, T.~Gutsche, S.~Kovalenko and V.~E.~Lyubovitskij,
  Phys.\ Rev.\  D {\bf 76}, 014003 (2007); 
  A.~Faessler, T.~Gutsche, V.~E.~Lyubovitskij and Y.~L.~Ma,
  Phys.\ Rev.\  D {\bf 76}, 114008 (2007); 
  Y.~Dong, A.~Faessler, T.~Gutsche and V.~E.~Lyubovitskij,
  Phys.\ Rev.\ D {\bf 77}, 094013 (2008). 
\bibitem{Mandelstam:1962mi}
 S.~Mandelstam,
 Ann. Phys.\  {\bf 19}, 1 (1962);
 J.~Terning,
 Phys.\ Rev.\ D {\bf 44}, 887 (1991).
\bibitem{Mergell:1995bf}
  P.~Mergell, U.~G.~Meissner and D.~Drechsel,
  Nucl.\ Phys.\  A {\bf 596}, 367 (1996). 
\bibitem{Kelly:2004hm}
  J.~J.~Kelly,
  Phys.\ Rev.\  C {\bf 70}, 068202 (2004).
\bibitem{TomasiGustafsson:2005ni}
  E.~Tomasi-Gustafsson, G.~I.~Gakh and C.~Adamuscin,
  Phys.\ Rev.\  C {\bf 73}, 045204 (2006). 
\bibitem{Abbott:2000ak}
  D.~Abbott {\it et al.}  (JLAB t(20) Collaboration),
  Eur.\ Phys.\ J.\  A {\bf 7}, 421 (2000). 
\bibitem{Kobushkin:1994ed}
  A.~P.~Kobushkin and A.~I.~Syamtomov,
  Phys.\ Atom.\ Nucl.\  {\bf 58}, 1477 (1995)
  [Yad.\ Fiz.\  {\bf 58}, 1565 (1995)]. 
\bibitem{Sick}I.~Sick and D.~Trautmann, 
Nucl. \ Phys. \ A {\bf 637}, 559 (1998). 
\bibitem{Abbott:2000fg}
  D.~Abbott {\it et al.}  [JLAB t(20) Collaboration],
  Phys.\ Rev.\ Lett.\  {\bf 84}, 5053 (2000). 
\bibitem{Bouwhuis:1998jj}
  M.~Bouwhuis {\it et al.},
  Phys.\ Rev.\ Lett.\  {\bf 82}, 3755 (1999). 
\bibitem{Garcon:1993vm}
  M.~Garcon {\it et al.},
  Phys.\ Rev.\  C {\bf 49}, 2516 (1994).
\bibitem{Gilman:1990vg}
  R.~A.~Gilman {\it et al.},
  Phys.\ Rev.\ Lett.\  {\bf 65}, 1733 (1990).
\bibitem{Schulze:1984ms}
  M.~E.~Schulze {\it et al.},
  Phys.\ Rev.\ Lett.\  {\bf 52}, 597 (1984).
\bibitem{Voitsekhovsky:1986xh}
  B.~B.~Voitsekhovsky {\it et al.},
  JETP Lett.\  {\bf 43}, 733 (1986)
  [Pisma Zh.\ Eksp.\ Teor.\ Fiz.\  {\bf 43}, 567 (1986)].
\bibitem{Nikolenko:2003zq} 
  D.~M.~Nikolenko {\it et al.},
  Phys.\ Rev.\ Lett.\  {\bf 90}, 072501 (2003).
\bibitem{Auffret:1985tg}
  S.~Auffret {\it et al.},
  Phys.\ Rev.\ Lett.\  {\bf 54}, 649 (1985).
\bibitem{Bosted:1989hy}
  P.~E.~Bosted {\it et al.}, 
  Phys.\ Rev.\  C {\bf 42}, 38 (1990). 
\bibitem{Cramer:1986kv}
  R.~Cramer {\it et al.},
  Z.\ Phys.\  C {\bf 29}, 513 (1985).
\bibitem{Abbott:1998sp}
  D.~Abbott {\it et al.}  [Jefferson Lab t(20) Collaboration],
  Phys.\ Rev.\ Lett.\  {\bf 82}, 1379 (1999). 
\bibitem{Platchkov:1989ch}
  S.~Platchkov {\it et al.}, 
  Nucl.\ Phys.\  A {\bf 510}, 740 (1990). 
\bibitem{Dmitriev:1985us}
V.~F.~Dmitriev {\it et al.}, 
Phys.\ Lett.\  B {\bf 157}, 143 (1985).
\bibitem{FerroLuzzi:1996dg}
  M.~Ferro-Luzzi {\it et al.},
  Phys.\ Rev.\ Lett.\  {\bf 77}, 2630 (1996).
\bibitem{PEPTS} 
M.~K.~Jones et al. [Jefferson Lab. Hall A Collaboration], 
Phys. Rev. \ Lett. {\bf 84}, 1398 (2000); 
O.~Gayou et al., [Jefferson Lab Hall A Collaboration], 
Phys. \ Rev. \ Lett. {\bf 88}, 092301 (2002).  
\bibitem{TPE} 
P.~A.~M.~Guichon and M.~Vanderhaeghen, 
Phys. \ Rev. \ Lett. {\bf 91}, 142303 (2003);  
P.~G.~Blunden, W.~Melnitchouk and J.~A.~Tjon,
Phys. \ Rev. \ Lett. {\bf 91}, 142304 (2003);
Y.~C.~Chen, A.~Afanasev, S.~J.~Brodsky, C.~E.~Carlson and
M.~Vanderhaeghen, Phys. \ Rev. \ Lett. {\bf 93}, 122301 (2004); 
A.~V.~Afanasev, S.~J.~Brodsky, C.~E.~Carlson, Y.~C. Chen and 
M.~Vanderhaeghen, Phys. \ Rev. \ D {\bf 72}, 013008 (2005); 
J.~Arrington, Phys. \ Rev. \ C {\bf 68}, 034325 (2003);
J.~Arrington, Phys. \ Rev. \ C {\bf 69}, 022201(R) (2004). 
\bibitem{Dong:2006wm}
  Y.~B.~Dong, C.~W.~Kao, S.~N.~Yang and Y.~C.~Chen,
  Phys.\ Rev.\  C {\bf 74}, 064006 (2006). 
\end{thebibliography}
\end{document}